# Intervention strategies for misinformation sharing on social media: A bibliometric analysis

Juanita Zainudin[1,2], Member, IEEE, Nazlena Mohamad Ali[1], Alan F. Smeaton[3], Fellow, IEEE and Mohamad Taha Ijab[1]
[1]Institute of Visual Informatics (IVI), Universiti Kebangsaan Malaysia, Bangi 43600, Malaysia
[2]Faculty of Computing and Multimedia, Universiti Poly-Tech Malaysia, Cheras 56100, Malaysia
[3]Insight Centre for Data Analytics, Dublin City University, Dublin 9, D09 Ireland

This work was supported by the Universiti Kebangsaan Malaysia research grant under Grant GPK-4IR-2020-019.

**ABSTRACT** Widely distributed misinformation shared across social media channels is a pressing issue that poses a significant threat to many aspects of society's well-being. Inaccurate shared information causes confusion, can adversely affect mental health, and can lead to mis-informed decision-making. Therefore, it is important to implement proactive measures to intervene and curb the spread of misinformation where possible. This has prompted scholars to investigate a variety of intervention strategies for misinformation sharing on social media. This study explores the typology of intervention strategies for addressing misinformation sharing on social media, identifying 4 important clusters – cognition-based, automated-based, information-based, and hybrid-based. The literature selection process utilized the PRISMA method to ensure a systematic and comprehensive analysis of relevant literature while maintaining transparency and reproducibility. A total of 139 articles published from 2013-2023 were then analyzed. Meanwhile, bibliometric analyses were conducted using performance analysis and science mapping techniques for the typology development. A comparative analysis of the typology was conducted to reveal patterns and evolution in the field. This provides valuable insights for both theory and practical applications. Overall, the study concludes that scholarly contributions to scientific research and publication help to address research gaps and expand knowledge in this field. Understanding the evolution of intervention strategies for misinformation sharing on social media can support future research that contributes to the development of more effective and sustainable solutions to this persistent problem.

## I. INTRODUCTION

Communication technology has made it easier for people to share information. The number of social media users is increasing every day. According to Statista [1], there are 5.18 billion Internet users worldwide, while social media usage has reached 4.8 billion users. and this number is expected to grow. There are many social media platforms available including Facebook, Instagram, Reddit, TikTok, Twitter, WeChat, Weibo, WhatsApp, and YouTube. Over the last 10-15 years, social media has essentially changed how we seek and share information. Social media channels provide an excellent platform for sharing news and accessing information around the world without any technical barriers or difficulties. Prior research by Valecha et al. [2] found that social networks have increased the connection and communication between social media users regardless of cost and distance. On the contrary, these platforms can and do turn into channels for the dissemination of misinformation.

There is often confusion about the sources of misinformation, which makes it difficult to identify it. Misinformation can appear on social media in several forms such as misleading, parody, imposter, false context, and manipulated content [3]. Thus, sharing misinformation worsens its impact and can increase anxiety among individuals [4]. It generates misperceptions among others that influence human decision-making in various domains including health, politics, and religion, which may disrupt societal harmony [5]. Misinformation in the health field can have a serious negative effect on mental health, increasing anxiety and depressive symptoms [6], [7]. In addition, people's trust in the credibility of news sources has diminished due to the extensive transmission of misleading information. For example, anti-vaccination groups have chosen not to get the COVID-19 vaccine due to misinformation, which has had an impact on public health [8], [9]. Meanwhile, in the political domain, misinformation has led voters to make unwise decisions during elections. For instance, the 2016 US presidential election is acknowledged to have contributed to the dissemination of misinformation on social media that somehow has an impact on voters' decisions [10], [11]. In the religious context, misinformation about religion may also pose a threat to society. One such instance is the propagation of false information about Islam that gave rise to Islamophobia [12]. Therefore, implementing intervention strategies to combat the spread of misinformation on social media is crucial due to its harmful effects on individual mental health and overall societal well-being, making research on these strategies essential.

There has been a significant increase in the range and volume of research on misinformation topics since 2020 because of the emergence of the COVID-19 pandemic in 2019. According to Patra et al. [13, p. 628], "the current pandemic on COVID-19 as a subject of study also marginally contributed to the world of fake information in 2020". The rise in misinformation studies also highlights the need for better education and awareness regarding the dangers of misinformation and the importance of using only reliable sources of information including the way to countermeasure the issues that misinformation causes. However, the area of study on intervention strategies against misinformation sharing is relatively new and evolving with rapid developments in the field. For instance, social media platforms and their associated agencies have put in place several regulations to control the spread of fake news. However, it is not enough to rely solely on social media regulation to control the spread of fake news. It is imperative to manage fake news comprehensively, which requires international cooperation [14].

There are several reasons for studying literature on this topic. Essentially, this literature review helps researchers recognize current research trends, allowing them to focus on areas that need exploration. The study's tendencies can also be used to identify research gaps. This guarantees the new study's applicability, significance, and impactful on society. Additionally, it is crucial to comparably analyze significant countries, publications, and articles in this area. By carefully examining these variables, we can learn important lessons about how to counteract misinformation and create strong solutions to deal with this widespread problem that could be applied as a global norm. Lastly, it's important to look at evolving themes and typologies of misinformation-sharing intervention approach issues. Understanding the dynamic nature of misinformation assists in the development and improvement of focused interventions, the promotion of disciplinary collaboration, the guiding of policy, and the identification of future research requirements. Adopting this strategy is crucial for effectively addressing the complex and dynamic challenges of misinformation in the current digital environment. The following are the research questions for this study.

1. RQ1: What are the current trends in intervention strategies for misinformation-sharing studies from 2013 to 2023?
2. RQ2: What are the most influential countries, journals, and articles for studying intervention strategies against misinformation-sharing topics?
3. RQ3: What are the evolving themes and typology for studying intervention strategies against misinformation-sharing topics?

This review paper consists of 7 sections. Following this section is Section 2 literature review which delves into the topic of intervention strategies against misinformation sharing on social media. Section 3 describes the methodology and process used in the review. Section 4 presents the results, including bibliometric analysis, theme clustering generation, and comparative analysis of the conclusions reached. Section 5 presents the discussion on the research objective, and Section 6 outlines the limitations and challenges faced. Finally, Section 7 presents the conclusion.

**II. LITERATURE REVIEW**

*A. INTERVENTION STRATEGIES AGAINST MISINFORMATION SHARING*

There are many ways that can be used as intervention strategies to mitigate the spread of misinformation on social media which can involve all the significant practitioners; individuals, technology platforms, and governments [15]. Firstly, individual-level intervention strategies encompass self-assessment of the content received by verifying its credibility against reputable fact-checking websites such as Snopes or PolitiFact but this requires effort on the part of the individual. Educating individuals and learning about how to identify reliable information sources can help people make informed judgments. In addition, self-verification must be supported by attentive-based design to intervene users to think before they decide to share misinformation on social media [16], [17]. The use of techniques such as boosting, false tags, nudging, warning, and visuals were expected able to intervene with user attention and trust before they decide to share [18], [19], [20], [21], [22], [23].

Secondly, to tackle the problem of fake news on social media platforms, platform-level intervention is desirable. This can involve using algorithmic detection techniques like crowdsourcing and third-party fact-checking, which rely on machine learning to minimize the need for human intervention and ensure high-quality performance by artificial intelligence (AI) techniques. These interventions can identify fake accounts and misinformation and apply platform-level filtering to counter the spread of fake news. However, relying solely on algorithms or bots to detect misinformation can result in inaccurate decisions [24]. In addition, Hamed et al. [25, p. 379] noted that "the accuracy of detection models is still notably poor". Therefore, technology-driven approaches should be complemented by human-based and crowd-sourced techniques to raise awareness of misinformation on social media platforms [15]. For example, social media platforms like YouTube and Facebook are equipped with advanced AI tools and skilled workforces to design and implement solutions to prevent fake news [26], [27]. In this regard, a crowdsourcing technique is required which uses feedback from users and third-party fact-checking to classify and flag fake news or false tags before they are shared. A related way to mitigate the unfavorable impact of fake news on social media is by enacting a platform-level policy. According to Papanastasiou [28], the effectiveness of the platform's policy in combating fake news depends on the prevalence of such news in the environment.

Finally, the process of government-level or other regulatory intervention entails the establishment of regulations and policies tailored to the unique concerns and issues of each country. This can prove to be a daunting task for nations with sizable populations, such as China and India, as noted by Rodrigues et al. [29]. It is important to note that certain government policies may inadvertently restrict freedom of expression, a concern highlighted by Vese [30]. To safeguard this fundamental right, it is crucial to foster an open dialogue and address any contentious policies to prevent the handling of fake news from undermining it.

### *B. UNDERSTANDING THE PRACTICAL IMPLICATIONS OF INTERVENTION STRATEGIES AGAINST MISINFORMATION SHARING*

The rationale of this study can be proven by outlining the practical implications of understanding intervention strategies against misinformation. Malaysia was one of the first countries to introduce and enforce laws related to fake news [31]. A prime example is the enforcement of Malaysia's Anti-Fake News Act 2018 (AFNA) [Act 803]. AFNA was enacted by the minister in the Prime Minister's Department of Malaysia on 11 April 2018, as the effective date of the Act [32]. The Act defined fake news broadly, including any news, information, data, and reports that are wholly or partly false [33]. The Act imposed severe penalties, including fines of up to RM 500,000 and imprisonment of up to six years for those found guilty which also applies to offenders outside Malaysia, including non-nationals, if Malaysia or a Malaysian citizen are affected [34]. The Act aimed to tackle the dissemination of misinformation that could impact public order and national security. The increase in fake news, especially during the 2018 general election in Malaysia, prompted the government to enact the Anti-Fake News Act (AFNA) as a measure to address these concerns [35], [36].

The impact before and after the implementation of the Act is notably significant in controlling the spread of misinformation. This was particularly evident during the 2018 general election campaign season, when misinformation about political figures and social issues was spread, causing confusion and a loss of trust in the media and government. For example, AFNA has led to the first case of a Danish national being jailed and fined for spreading fake news about a distress call response and an alleged assassination [36]. This has cautioned the public and led them to be more cautious about sharing uncertain news. However, AFNA has received strong criticism from domestic and international observers [34], [35]. The Act was too broad and seized the freedom of public speech, indirectly silencing political dissent [35]

AFNA was repealed in October 2019 after the change of Federal government, in response to public criticism [37]. The new government claimed that the AFNA undermined civil rights and that the regulations already in place were adequate to address misinformation [36]. However, the repeal did not solve the misinformation issues, which continued to plague the public during the COVID-19 pandemic and presented even greater challenges. Consequently, the Malaysian government and various stakeholders made concerted efforts to promote media literacy, fact-checking, and enforcing existing laws to counter misinformation, all without perceived overreach of AFNA [38], [39], [40].

In conclusion, the AFNA represents the practical implication of a real-life case where intervention strategies have been used to tackle the spread of misinformation. While these strategies have had some success in controlling the circulation of fake news, they also raise concerns about freedom of speech. Therefore, further research is needed to develop more effective intervention strategies that can improve upon the existing methods and better serve the nation.

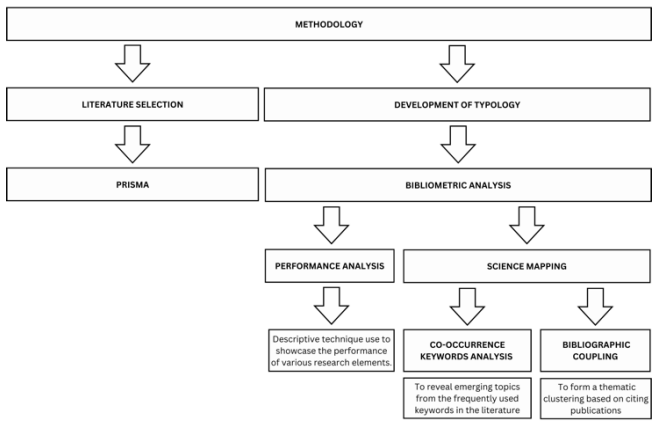

**FIGURE 1.** Methodology process.

## III. METHODS

The methodology includes literature selection and the development of typology for this study. The literature selection process utilized the PRISMA method, while the development of typology involved bibliometric analysis using performance analysis and science mapping techniques as illustrated in Fig. 1.

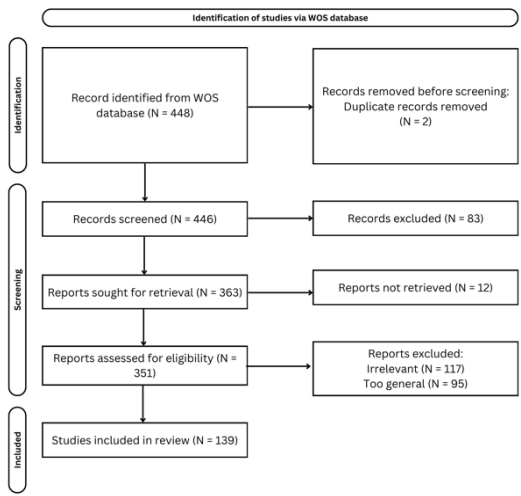

**FIGURE 2.** PRISMA flow diagram for literature selection [43].

### A. LITERATURE SELECTION

The PRISMA (Preferred Reporting Items for Systematic Reviews and Meta-Analyses) method was systematically employed during the literature selection process to identify, screen, and classify relevant scholarly works on intervention strategies for misinformation-sharing topics. Fig. 2 illustrates the steps of the selection process, including identification, screening, and inclusion. This method is a systematic approach for conducting literature reviews in a structured and transparent manner, including bibliometric reviews, as described in numerous research papers [41], [42]. To ensure thorough coverage of the relevant research, the review process is guided by the use of flowcharts and item lists [43].

For this study, articles were extracted from the Web of Science (WOS) database within a range of 10 years (2013-2023). The study chooses the WOS database for its comprehensive coverage of scientific literature, making it a reliable source for bibliometric analysis with high-quality content and reputable scientific journal indexing, ensuring credible and accurate data retrieval for analysis [44], [45]. The search terms were divided into three parts (see Table I):

TABLE I
SEARCH CRITERIA

| Search query | | Results |
|---|---|---|
| Search year | 2013 to 2023 | |
| Search terms | Part 1: (misinform* OR "wrong inform*" OR disinform* OR "misleading inform*" OR "fake news") | 448 |
| | AND | |
| | Part 2: ("intervention technique*" OR "intervention strateg*" OR intervention*) | |
| | AND | |
| | Part 3: ("social media*" OR Facebook* OR Instagram* OR Reddit* OR TikTok* OR Twitter* OR WeChat* OR Weibo* OR Whatsapp* OR YouTube) | |
| Document type | Article | 363 |

1. The first part consists of terms included in or related to the umbrella term of "misinformation". Keywords like wrong information, disinformation, misleading, and fake news were also used interchangeably to represent misinformation. Since the term misinformation is used in different ways by scholars such as misinformation or misinforming, all these options were considered as keywords in the article searching process.
2. The second part contains a list of all related synonyms for intervention strategies and related concepts. The term "intervention strategy" focuses on concrete actions, plans, and policies to address misinformation. It narrows down the literature to works directly concerned with interventions, making the review process more manageable and ensuring applicability. Literature reviews may be dominated by studies that only describe a problem without offering solutions, therefore using an "intervention strategy" can bias the search toward studies that propose, test, or analyze specific strategies for addressing the issue.

3. The third part highlights social media by including relevant terms that relate to it such as Facebook, Instagram, Reddit, TikTok, Twitter, WeChat, Weibo, WhatsApp, and YouTube. Different platforms may employ or require different intervention strategies. By including a wide range of social media platforms, the review can assess a broader spectrum of strategies, from algorithmic interventions to user education and policy changes.

After conducting a keyword search on the WOS database, 139 out of 448 initial records met the eligibility criteria for inclusion in the bibliometric analysis. Duplicate articles were removed during the filtering process reducing the number of records to 446. Duplicate articles refer to identical articles in content that may have been collected from multiple sources or indexed multiple times. The data was then screened based on the document type "Article", which returned 363 articles. A total of 363 publications were sought to be retrieved, of which 12 were not able to be retrieved. This left us with 351 records that were assessed for eligibility. A thorough examination of the abstracts led to the identification of 139 publications that met the criteria and were related and relevant to the search. During the filtering process, 60.4% of the articles were excluded as their topic was irrelevant (117 records) or too general (95 records). Records were excluded for several reasons. Firstly, the topics were irrelevant as they focused on issues unrelated to intervention strategies for misinformation-sharing behavior, such as vaccine hesitancy and vaccine-related issues. Additionally, studies that were too broad, like articles concentrating on human attitudes toward fake news, were also excluded. We also omitted literature primarily focused on medical or psychological domains that address health and psychological issues rather than intervention strategies.

### B. DEVELOPMENT OF TYPOLOGY

A bibliometric analysis was utilized to develop a typology based on a theme. A bibliometric analysis using performance analysis and science mapping techniques was used to summarize and outline the recent intellectual structure and emerging trends related to the topic [46].

Performance analysis is a descriptive technique used to showcase the performance of various research elements such as countries, journals, and articles within a specific field. Its purpose is to identify current trends and the most influential countries, journals, and articles for studying the given topic.

On the other hand, science mapping involves analyzing the interconnections between research elements using methods like co-occurrence keyword analysis and bibliographic coupling. Co-occurrence keyword analysis aims to reveal emerging topics from the frequently used keywords in the literature, while bibliographic coupling helps in forming thematic clusters based on citing publications. VOS viewer software was used to analyze and visualize the meta-data to create maps such as co-occurrence and bibliographic coupling networks. The software is intended primarily for bibliometric networks which can create, visualize, and explore maps [47]. Data was visualized in tables and graphically to show meaningful information according to the year of publication, research topics, research area, countries, most productive journals, and articles with high citations.

### IV. RESULTS

We now provide an in-depth analysis result of the bibliometric findings from 139 articles screened using PRISMA guidelines which will contribute to the identification of theme clustering and typology in the topic of intervention strategies against misinformation sharing and answering our research questions.

#### A. YEAR OF PUBLICATION

There has been a steady increase in the number of publications related to intervention strategies against misinformation sharing on social media topics since 2019. In 2015 and 2018, only 2 publications were recorded respectively, but the number rose to 11 in 2019 and has continued to climb ever since. The number increased to 28 in 2021 and 48 in 2022. For this review paper, data were collected from publications published in 2013 until the first half of 2023, witnessing 32 publications in 2023. Refer to Fig. 3 for the trend in publications related to the intervention strategies for misinformation sharing on social media topic.

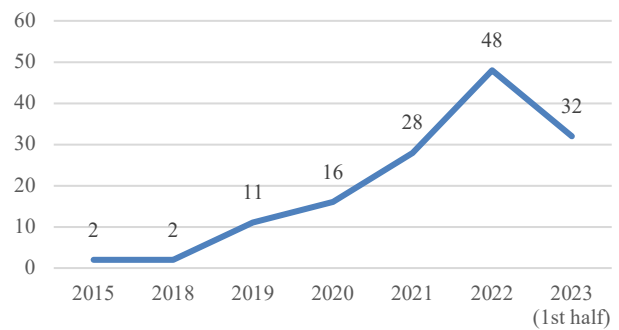

FIGURE 3. Yearly trend of published articles (2013 – 2023).

#### B. RESEARCH TOPICS TREND

This subsection focuses on the trending topics related to "intervention strategies for misinformation sharing" over the years, showing the evolution of these topics from 2013-2023. The analysis involved grouping articles by year and analyzing their abstracts manually.

In the last decade, research has been done to combat misinformation across various domains including psychology, communication, and computer science. The psychology and communication fields emphasize the impact of misinformation, the use of knowledge-based literacy for intervention strategies and sharing behavior from psychological and communication perspectives. This contrasts with the computer science field where treatment of the topic mostly focuses on evaluating the proposed model, tools, and machine learning in mitigating fake news. The analysis findings demonstrate a significant transition over the years, with topics shifting from knowledge-based (KB) to technology-based (TB) as shown in Table II.

TABLE II
TOPICS TREND OVER YEARS

| Intervention Trend | Scholar Topics | Range of Year |
|---|---|---|
| Knowledge-based | <ul><li>issue of misinformation sharing</li><li>information literacy</li><li>pledges as an intervention</li><li>regulating fake news</li></ul> | 2015- 2018 |
| Knowledge-based<br><br>Technology-based | <ul><li>credibility perceptions of false news stories on social media</li><li>fake news data management and mining</li><li>guideline for social media user on public health agenda</li><li>the effects of visual anchors and strategy cues</li><li>diffusion of pro- and anti-false information tweets</li><li>health guidance</li><li>media literacy</li><li>fake news detection</li></ul> | 2019-2020 |
| Technology-based | <ul><li>framework for infodemiology</li><li>attention-based system</li><li>storytelling simulation software</li><li>fake news detection and machine learning</li><li>fact-checking techniques</li><li>artificial intelligence (AI)</li></ul> | 2021-2023 |

Between 2015 and 2018, publications addressed topics related to the issue of misinformation sharing, and studies primarily centered on evaluating and reviewing KB intervention strategies. These interventions included efforts to enhance information literacy among consumers by understanding the reasons behind misinformation sharing [48], examining the effectiveness of pledges as an intervention to help address the misinformation-sharing problem [49], and regulating fake news and other online advertising [50]. These studies were all aimed at mitigating the spread of misinformation by educating individuals on information literacy and regulations implemented by governing bodies.

Further years have seen publications that cover a wider range of aspects in various fields with different points of view on their findings. In 2019 and 2020, some of the topics focused on the impact of user engagement on credibility perceptions of false news stories on social media [51], fake news data management and mining [52], guideline for social media user on public health agenda [53], conducting experiments to analyze the effects of visual anchors and strategy cues [22], diffusion of pro- and anti-false information tweets [54], quantifying COVID-19 content among online opponents of establishment health guidance [55], analyzing the level of media literacy to process fake news on social media [56], and conducting a survey to review and evaluate methods that can detect fake news [57]. In addition, the year 2020 found several noteworthy studies on COVID-19 conducted by previous researchers [29], [55], [58]. Over the years 2021-2023, there has been a significant increase in the number of topics related to the pandemic [8], [24], [29], [59], [60], [61]. During these years (2019-2020), scholars have noticed a shift in interest toward topics related to intervention strategies against misinformation sharing, from KB approaches to TB solutions.

Recent analysis has shown that articles published between 2021 and 2023 have emphasized TB strategies like the development of frameworks or tools for intervention strategies. For instance, a study by Scales et al. [62] has emphasized the theoretical framework for conducting Motivational Interviewing (MI)-based infodemiology interventions among digital communities. Researchers have also explored attention-based systems designed to intervene in user attention against sharing misinformation [24], as well as the use of storytelling simulation software to combat misinformation on social media [63]. Moreover, some topics studies on automatic detection techniques using graph-based methods and machine learning, to address the challenge of detecting misinformation and fake news, including experimenting with the Temporal Graph Neural Network (TGNN) model to highlight the importance of temporal interaction information in detecting fake news [64], introduce the COVAXLIES dataset and a method called "Misinformation Detection as Graph-Link Prediction" for detecting misinformation about COVID-19 vaccines [65], and examining participants sharing behavior and detection ability [11]. During that same time frame, topics like the use of fact-checking techniques to combat misinformation [66], [67], [68] were also explored. This included the implementation of warning labels informing users that the content has been disputed by fact-checkers [19]. Topics related to AI in combating misinformation sharing were also examined [26], [69].

In conclusion, more publications show the importance of this topic to researchers. Analysis shows a shift from KB strategies to TB interventions, including automatic detection, machine learning, fact-checking, and AI, to combat the spread of misinformation.

### C. POPULAR RESEARCH AREAS

The 139 articles retrieved through the WOS database were categorized into 29 research areas. The analysis specifically focused on the "Research Areas" field, with each article being associated with one or more categories. The categories were quantified using a spreadsheet and then ranked in descending order to identify the most significant research areas. Table III illustrates the top 10 areas that generated the highest number of articles among the 139. It is worth noting that numerous fields have exhibited a strong interest in developing, modeling, evaluating, and experimenting with strategies aimed at curbing the propagation of misinformation on social media platforms. Computer Science (26.6%) dominates the list of research areas, where researchers employed diverse techniques including tools, models, and detection mechanisms to combat misinformation. Communication (18.0%) and Information Science & Library Science (15.8%) have conducted surveys and experimental studies on various aspects of intervention strategies, including models, literacy, regulation, policy, and theories. Psychology (12.2%) ranks as the fourth most popular area, where researchers have primarily focused on exploring psychological factors that influence misinformation-sharing behavior, such as user motivation, self-efficacy, cognitive reflection, and epistemic belief.

The remaining articles, which contribute less than 10% of the total articles, have been grouped under various categories, such as Science & Technology - Other Topics, Business & Economics, Health Care Sciences & Services, Medical Informatics, Public, Environmental & Occupational Health, Government & Law, and others. In summary, the importance of this topic has been recognized across multiple areas, and the number of publications has been progressively increasing. Some articles have even covered more than one area of study. Thus, it is evident that the focus and scope of studies on intervention strategies against misinformation-sharing on social media vary significantly. This outcome can be a future direction for researchers to choose a research area and topic related to intervention strategies for misinformation sharing on social media.

TABLE III
TOP 10 RESEARCH AREAS FOR THE "INTERVENTION STRATEGIES FOR MISINFORMATION SHARING" TOPIC

| *R | Research Area | No. of Articles | % of 139 |
|---|---|---|---|
| 1. | Computer Science | 37 | 26.6% |
| 2. | Communication | 25 | 18.0% |
| 3. | Information Science & Library Science | 22 | 15.8% |
| 4. | Psychology | 17 | 12.2% |
| 5. | Science & Technology - Other Topics | 13 | 9.4% |
| 6. | Business & Economics | 12 | 8.6% |
| 7. | Health Care Sciences & Services | 12 | 8.6% |
| 8. | Medical Informatics | 10 | 7.2% |
| 9. | Public, Environmental & Occupational Health | 10 | 7.2% |
| 10. | Government & Law | 7 | 5.0% |

*R: Research area ranking.

### D. MOST PRODUCTIVE COUNTRIES

This subsection presented countries' contributions to publications related to intervention strategies against misinformation-sharing topics. Upon analyzing the data, it was discovered that the authors of the 139 published articles were from 35 different countries. Here are the Top 10 countries that have published articles in relation to intervention strategies for misinformation on social media in Table IV. The USA had the highest number of published articles with 68, contributing to 48.9% of the total publications. Following the USA, UK and the People's Republic of China were the next highest contributors, each publishing 18 articles (12.9%). Nigeria and Canada were then listed, each with 10 (7.2%) articles respectively. Germany, and Malaysia each produced 8 articles (5.8%) respectively, while Australia produced 6 articles (4.3%). Italy and the Netherlands produced a total of 5 articles (3.6%). The analysis shows that ranking by number of publications of a country does not reflect the total number of citations. For instance, UK was rated at second ranking of contributing more publications, but Canada contributed second for the highest number of citations.

The analysis revealed that the USA has become the top country in publishing articles related to this topic, likely due to its major concern about the spread of fake news online. This is in line with findings by Akram et al. [70] and Wang et al. [71] who conducted a bibliometric analysis on misinformation and reported that the USA appears to be the most influential country with its more significant role in advancing misinformation research.

TABLE IV
MOST PRODUCTIVE COUNTRIES FOR THE "INTERVENTION STRATEGIES FOR MISINFORMATION SHARING" TOPIC

| *R | Country | Total Publications | Total Citations |
|---|---|---|---|
| 1 | USA | 68 | 3041 |
| 2 | UK | 18 | 686 |
| 3 | People's Republic of China | 18 | 238 |
| 4 | Nigeria | 10 | 80 |
| 5 | Canada | 10 | 1275 |
| 6 | Germany | 8 | 173 |
| 7 | Malaysia | 8 | 70 |
| 8 | Australia | 6 | 382 |
| 9 | Italy | 5 | 96 |
| 10 | Netherlands | 5 | 13 |

*R: Country ranking.

### E. MOST PRODUCTIVE JOURNALS

A study analyzing journals has revealed that between 2013 and 2023, 98 journals published articles on topics related to intervention strategies against misinformation sharing on social media. The research was conducted on the sample of 139 articles, and the top 20 journals were ranked based on the number of publications, followed by citation count. The Journal of Medical Internet Research was found to be the most productive in this field, with 7 publications and 67 citations. Scientific Reports followed closely behind with 3 articles and 44 citations, a higher number of citations, putting it in second place. Likewise, journals such as Social Media + Society, Internet Research, Information Systems Frontiers, New Media & Society, Digital Journalism, and Computers in Human Behavior also yielded 3 articles each.

Based on the analysis, it was found that research articles related to intervention strategies for combating misinformation sharing on social media were published in reputable journals with high impact factors. The significant number of citations these articles received indicates the crucial nature of this topic. Table V illustrates the top 20 of the most productive journals for the topic of intervention strategies for misinformation sharing on social media. Data analysis shows a gap in articles related to intervention strategies against misinformation in most journals. Even the top-performing journal has only published 7 articles on the subject. This highlights the importance of the topic for researchers submitting related articles.

TABLE V
MOST PRODUCTIVE JOURNALS FOR THE "INTERVENTION STRATEGIES FOR MISINFORMATION SHARING" TOPIC

| *R | Journal | Total # Publications | Total # Citations | Impact Factor (2022) | 5 Years Impact Factor (2017-2022) |
|---|---|---|---|---|---|
| 1 | JOURNAL OF MEDICAL INTERNET RESEARCH | 7 | 69 | 7.4 | 7.6 |
| 2 | SCIENTIFIC REPORTS | 3 | 44 | 4.6 | 4.9 |
| 3 | SOCIAL MEDIA + SOCIETY | 3 | 16 | 5.2 | 6 |
| 4 | INTERNET RESEARCH | 3 | 16 | 5.9 | 7.9 |
| 5 | INFORMATION SYSTEMS FRONTIERS | 3 | 16 | 5.9 | 6 |
| 6 | NEW MEDIA & SOCIETY | 3 | 14 | 5 | 6.9 |
| 7 | DIGITAL JOURNALISM | 3 | 6 | 5.4 | 6.4 |
| 8 | COMPUTERS IN HUMAN BEHAVIOR | 3 | 3 | 9.9 | 10.2 |
| 9 | INFORMATION AND LEARNING SCIENCES | 2 | 0 | 3.4 | 2.3 |
| 10 | PSYCHOLOGICAL SCIENCE | 2 | 711 | 8.2 | 8.4 |
| 11 | HEALTH INFORMATICS JOURNAL | 2 | 509 | 3 | 3 |
| 12 | PLOS ONE | 2 | 211 | 3.7 | 3.8 |
| 13 | MANAGEMENT SCIENCE | 2 | 170 | 5.4 | 7.1 |
| 14 | JOURNAL OF ACADEMIC LIBRARIANSHIP | 2 | 109 | 2.6 | 2.1 |
| 15 | NATURE HUMAN BEHAVIOUR | 2 | 79 | 29.9 | 23.8 |
| 16 | ONLINE INFORMATION REVIEW | 2 | 47 | 3.1 | 3.3 |
| 17 | IEEE ACCESS | 2 | 47 | 3.9 | 4.1 |
| 18 | INFORMATION PROCESSING & MANAGEMENT | 2 | 36 | 8.6 | 8.2 |
| 19 | ANNALS OF THE AMERICAN ACADEMY OF POLITICAL AND SOCIAL SCIENCE | 2 | 17 | 2.8 | 3.2 |
| 20 | JOURNAL OF MANAGEMENT INFORMATION SYSTEMS | 2 | 16 | 7.7 | 11.4 |

*R: Journal ranking.

### F. HIGH CITATION ARTICLES

In this subsection, the analysis focused on the most frequently cited articles from the selection of 139 publications. Table VI displays the top 10 articles with the highest number of citations. The most-cited article was published in the Psychological Science journal by Pennycook et al. [58], with 667 citations. The research in that paper focused on a nudging intervention strategy to encourage individuals to consider accuracy before sharing on social media. In the study, a survey of US adults conducted online revealed that nudging them to think about accuracy improved their social media decision-making. As found in the previous subsection on productive journal analysis, Psychological Science was ranked as the top journal in terms of the number of citations. This shows that Pennycook et al. [58]'s article contributed significantly to this ranking.

The article with the second-highest number of citations was published in the Health Informatics Journal by Madathil et al. [72], with a total of 479 citations. This article contributed to the overall number of citations in the Health Informatics Journal, making it one of the most highly cited journals in this field. The paper conducted a systematic review of various works related to healthcare information on YouTube, emphasizing the need for interventions that empower consumers to make informed decisions. Researchers searching for healthcare-related topics often cite this review paper, which was published in 2015.

The next-highest article was a study by Zhou et al. [57] which was published in ACM COMPUTING SURVEYS and had accumulated 284 citations. This survey paper reviewed and evaluated methods that can detect fake news from four perspectives: the false knowledge it carries, its writing style, its propagation patterns, and the credibility of its source.

Following the above were more highly-cited articles about intervention strategies against misinformation sharing on social media: (1) Pennycook et al. [16], published in the NATURE journal, with 218 citations, which proposes attention-based interventions to counter misinformation on social media, (2) Islam et al. [9], published in the PLOS ONE journal, with 185 citations, who study COVID-19 vaccine rumours and conspiracy theories and suggest interventions to manage misinformation and increase vaccine acceptance, (3) Sharma et al. [73], published in the ACM Transactions journal, with 171 citations, which surveys the technical challenges of fake news identification and mitigation and summarizes available datasets, (4) Pennycook et al. [20], published in Management Science journal with 138 citations, challenging theories of motivated reasoning and identifying a potential challenge for using warning tags to fight misinformation by demonstrating an implied truth effect where untagged false headlines are considered more accurate, (5) Guess et al. [74], published in Proceedings of The National Academy of Sciences of the United States of America with 130 citations, evaluating the effectiveness of an intervention model closely related to the world's largest media literacy campaign, which improved the ability to discern between mainstream and false news headlines, (6) Chen et al. [48], published in the Journal of Academic Librarianship with 106 citations, analyzing the root causes of misinformation dissemination and provided valuable guidance on how to improve information literacy intervention strategies, and (7) Walter et al. [75], published in the Health Communication Journal with 101 citations, conducting a study using a meta-analysis to evaluate the relative impact of social media interventions designed to correct health-related misinformation, in which theory-driven moderators help differentiate the effectiveness of social media interventions.

TABLE VI
TOP 10 RANKING OF THE MOST CITED ARTICLES

| *R | Article Title | Authors | Journal | Year | *TC |
|---|---|---|---|---|---|
| 1 | Fighting COVID-19 Misinformation on Social Media: Experimental Evidence for a Scalable Accuracy-Nudge Intervention | Pennycook, G; McPhetres, J; Zhang, YH; Lu, JG; Rand, DG | PSYCHOLOGICAL SCIENCE | 2020 | 667 |
| 2 | Healthcare information on YouTube: A systematic review | Madathil, KC; Rivera-Rodriguez, AJ; Greenstein, JS; Gramopadhye, AK | HEALTH INFORMATICS JOURNAL | 2015 | 479 |
| 3 | A Survey of Fake News: Fundamental Theories, Detection Methods, and Opportunities | Zhou, XY; Zafarani, R | ACM COMPUTING SURVEYS | 2020 | 284 |
| 4 | Shifting attention to accuracy can reduce misinformation online | Pennycook, G; Epstein, Z; Mosleh, M; Arechar, AA; Eckles, D; Rand, DG | NATURE | 2021 | 218 |
| 5 | COVID-19 vaccine rumors and conspiracy theories: The need for cognitive inoculation against misinformation to improve vaccine adherence | Islam, MS; Kamal, AHM; Kabir, A; Southern, DL; Khan, SH; Hasan, SMM; Sarkar, T; Sharmin, S; Das, S; Roy, T; Harun, MGD; Chughtai, AA; Homaira, N; Seale, H | PLOS ONE | 2021 | 185 |
| 6 | Combating Fake News: A Survey on Identification and Mitigation Techniques | Sharma, K; Qian, F; Jiang, H; Ruchansky, N; Zhang, M; Liu, Y | ACM TRANSACTIONS ON INTELLIGENT SYSTEMS AND TECHNOLOGY | 2019 | 171 |
| 7 | The Implied Truth Effect: Attaching Warnings to a Subset of Fake News Headlines Increases Perceived Accuracy of Headlines Without Warnings | Pennycook, G; Bear, A; Collins, ET; Rand, DG | MANAGEMENT SCIENCE | 2020 | 138 |
| 8 | A digital media literacy intervention increases discernment between mainstream and false news in the United States and India | Guess, AM; Lerner, M; Lyons, B; Montgomery, JM; Nyhan, B; Reifler, J; Sircar, N | PROCEEDINGS OF THE NATIONAL ACADEMY OF SCIENCES OF THE UNITED STATES OF AMERICA | 2020 | 130 |
| 9 | Why Students Share Misinformation on Social Media: Motivation, Gender, and Study-level Differences | Chen, XR; Sin, SCJ; Theng, YL; Lee, CS | JOURNAL OF ACADEMIC LIBRARIANSHIP | 2015 | 106 |
| 10 | Evaluating the Impact of Attempts to Correct Health Misinformation on Social Media: A Meta-Analysis | Walter, N; Brooks, JJ; Saucier, CJ; Suresh, S | HEALTH COMMUNICATION | 2021 | 101 |

*R: Article ranking.
*TC: Total citations.

### G. KEYWORD CO-OCCURRENCE ANALYSIS

A bibliometric analysis was conducted on the co-occurrence of author keywords, based on the selection of 139 papers presented in Table VII. The aim was to uncover the emerging topics from the frequently used keywords of the literature, thus answering the third research question. The final keyword analysis was based on author-provided keywords that occurred at least 5 times, resulting in a total of 30 keywords that met the threshold requirements out of the initial 614 keywords. The co-occurrence analysis also suggests that frequently appearing words are thematically related, while author-defined keywords that co-occur suggest spatially close themes [46].

TABLE VII
KEYWORD CO-OCCURRENCE

| *R | Keyword | Occurrences | Total Link Strength |
|---|---|---|---|
| 1 | social media | 72 | 204 |
| 2 | misinformation | 60 | 196 |
| 3 | fake news | 55 | 186 |
| 4 | information | 20 | 72 |
| 5 | covid-19 | 25 | 64 |
| 6 | disinformation | 16 | 57 |
| 7 | credibility | 10 | 38 |
| 8 | communication | 9 | 36 |
| 9 | fact-checking | 8 | 36 |
| 10 | media literacy | 8 | 36 |
| 11 | online | 9 | 35 |
| 12 | news | 11 | 34 |
| 13 | trust | 10 | 34 |
| 14 | twitter | 11 | 34 |
| 15 | media | 10 | 30 |
| 16 | model | 8 | 30 |
| 17 | false news | 6 | 29 |
| 18 | continued influence | 8 | 27 |
| 19 | facebook | 10 | 25 |
| 20 | bias | 6 | 23 |
| 21 | knowledge | 8 | 21 |
| 22 | behavior | 6 | 19 |
| 23 | deception | 5 | 19 |
| 24 | health information | 5 | 19 |
| 25 | literacy | 6 | 18 |
| 26 | web | 5 | 17 |
| 27 | digital literacy | 5 | 16 |
| 28 | health | 5 | 16 |
| 29 | inoculation theory | 5 | 16 |
| 30 | persuasion | 5 | 15 |

*R: Keyword ranking.

The presence of keywords like "social media", "Twitter", and "Facebook" depicts that the intervention strategies against misinformation sharing have been widely applied to studies on social media platforms. An examination of keywords reveals that diverse approaches are being implemented to combat the spread of misinformation on social media. These methods encompass the creation of a "model", promoting "digital literacy", raising awareness on "media literacy", and employing "fact-checking" techniques to flag misinformation. Furthermore, research on strategies to intervene against the dissemination of misinformation concentrates on leveraging theories like "persuasion" and "inoculation" to alleviate the issue of misinformation sharing "behavior".

### H. BIBLIOGRAPHIC COUPLING AND THEME CLUSTERING GENERATION

This subsection conducts a bibliographic coupling analysis from 139 selected documents. After keeping the minimum number of citations to 10, 48 documents for analysis as illustrated in Fig. 4. This analysis is performed to understand the evolving theme for intervention strategies in mitigating misinformation sharing on social media. Findings from the analysis have revealed that 4 clusters were identified.

*Cluster 1 (red colour): Cognition-based as an intervention strategy.*
The first cluster identified in red colour has 14 articles, with the theme of a "cognition-based" intervention strategy against misinformation sharing on social media. This theme provides works of literature that focus on nudging techniques (7 articles), media literacy (4 articles), and inoculation theory (2 articles). The article by Rodrigues et al. [29] is dropped because it does not belong to this cluster. The similarity provided by literature under this theme is, that all the studies focus on strategies to combat misinformation sharing on social media using a cognition-based approach.

Cognition refers to a strategy of educating people on media literacy and nudging people to be more careful, which focuses on the use of thinking abilities in mitigating misinformation sharing. Cognitive abilities can be nudged using tools such as warnings and reminders [16], [20], [21], [58], [76], [77], [78], can be educated on media literacy using campaigns and guidelines [56], [74], [79], [80], and can be immunized using inoculation theory [81], [82].

Cognitive tools as an intervention that uses warnings or reminders can nudge people's attention to accuracy when comes to decisions in sharing misinformation behavior. Pennycook et al. [16] have suggested that nudging individuals to think about accuracy is an effective way to improve their choices about what to share on social media.

Nevertheless, understanding media literacy can increase discernment between mainstream and false news among social media users. Prior research by Guess et al. [74] has shown that campaigns to promote media literacy, which provide tips on how to spot false news, can be an effective way to combat false or misleading news. This finding has significant real-world implications. It is recommended that people use these campaigns to help them learn to think critically about what they read or hear. They can help people see through the fake news.

Additionally, inoculation theory is another way that can prevent people from sharing misinformation on social media. According to the theory, people can build up resistance to persuasive messages, much like they can become immune to viruses [82]. This means that individuals can prepare themselves psychologically to resist such messages. For example, the use of the Bad News Game, as an inoculation-based intervention for media and information literacy, can protect against misinformation influence over time [81].

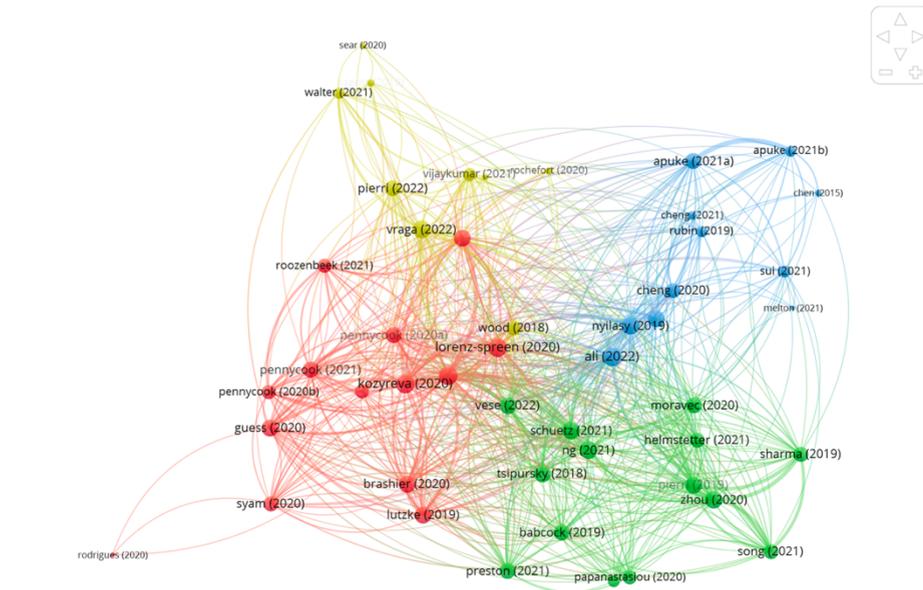

**FIGURE 4.** Bibliographic coupling of articles.

*Cluster 2 (green color): Automated-based as an intervention strategy.*
The second cluster consists of 14 articles related to "automated-based", which are highlighted in green. However, the article Albrecht et al. [83] is not related to this theme and is therefore excluded. The literature on this theme is diverse, with a focus on four main objectives: 1) identifying fake news through various means [54], [73], [84], [85], [86], 2) using algorithms to detect fake news [57], [64], [73], [84], 3) employing fact-checking approaches [49], [67], [87], and 4) developing policies and regulations for addressing fake news at the platform level [28], [30], [49], [88].

Identifying the types of fake news is a crucial step in understanding the pattern and classifying them into different categories. By doing so, we can come up with a comprehensive guideline to propose a mitigation solution. This will not only help to combat the spread of misinformation but also ensure that people have access to accurate information. For instance, researchers like [54], [84], [85], and [73] have identified and classified the criteria of misinformation and available datasets in their studies. Thus, the findings were useful for implementing misinformation detection intervention strategies like algorithmic detection and fact-checking websites.

Algorithmic detection relies on machine learning algorithms that analyze content features such as linguistic patterns or metadata to automatically flag potentially fake news articles for further review. A review study conducted by Zhou et al. [57] identified methods that detect fake news from four different perspectives. These perspectives include the false knowledge that fake news communicates, its writing style, its propagation patterns, and the credibility of its source.

Furthermore, fact-checking is another technique used to verify the accuracy and truthfulness of information. It involves the process of analyzing and evaluating claims or statements to determine their validity. Fact-checking can be performed by specialized agencies, as well as through alternative educational curricula and the involvement of professionals in information literacy. For example, PolitiFact, a fact-checking organization, has been diligently scrutinizing the statements made by politicians for accuracy since 2007 [87]. They evaluate the veracity of these statements and assign a rating based on their findings, thus providing the public with a comprehensive and reliable source of information. In this regard, researchers like Moravec et al. [87] and Schuetz et al. [67] have explored the usage of fact-checking techniques to combat misinformation sharing. Their studies have mainly focused on the importance of fact-checking techniques in controlling the spread of misinformation. Thus, it was undoubtful that the fact-checking technique had become an established way of detecting misinformation in social media.

Meanwhile, developing policies and regulations at the platform level is important to strengthen the usage of false flags or fact-checks. According to Ng et al. [88], implementing a forwarding restriction policy leads to less direct forwarding of fake news compared to truthful news. This is due to social tie theory, which explains that strong ties to fake news sources are prevented from spreading fake news, while weak ties are not affected. In this regard, the research demonstrates that implementing forwarding restriction policies can shorten the lifespan of fake news. Governments around the world, at both the international and European Union (EU) levels, are introducing legislative and administrative measures to control the spread of fake news on social media. However, these measures could also result in limitations on freedom of expression and increased censorship. In a study conducted by Vese [30], the negative implications of these measures were analyzed, and alternative regulatory approaches in public law were suggested. The study proposed self-regulation and empowering users as strategies to combat fake news and recommends implementing reliability ratings on social media platforms. In another study, Papanastasiou [28] has highlighted the importance of the platform's policy that affects how people learn about news. The study found that when there is very little fake news, the platform's policy is more effective if it makes it very clear that sharing fake news is bad. When there is a lot of fake news, the platform's policy is more effective if it offers small rewards for sharing news and big penalties for sharing fake news.

*Cluster 3 (blue color): Information-based as an intervention strategy.*
The third cluster, which is represented by a blue linked node, consists of 11 articles that focus on information-based studies as an intervention strategy against misinformation sharing. Most of the literature under this theme examines the "information-based" intervention strategy, construction of models, and explores the factors of sharing intention to combat the spreading of misinformation on social media. This theme can be divided into three categories: 1) information elements [48], [89], [90], [91], 2) constructing models [92], [93], [94], [95], and 3) sentiment analysis [96].

The content shared on social media has a significant impact on people's sharing behavior. Ali et al. [89] suggest that the features of the information shared on social media can influence people's decision-making process when it comes to sharing. Therefore, strategies that utilize information elements such as information cues, characteristics, and literacy can significantly affect people's sharing intentions. For instance, individuals who attend social media-based counseling have been shown to have a more positive perception of the COVID-19 vaccine [91]. Such counseling sessions can help educate people to become more information literate about social media-related issues. Additionally, understanding the motivation and characteristics of people who share misinformation on social media is critical in developing information literacy intervention strategies to reduce misinformation sharing. According to Chen et al. [48], one of the main reasons people share misinformation on social media is due to the information's perceived characteristics.

The second category in this theme includes studies that focus on constructing a model based on existing theories to predict the behavior of sharing misinformation. For example, prior studies [92], [93] have developed several theoretical models that predict the sharing of fake news during the COVID-19 pandemic. The findings have revealed that the abundance of information on COVID-19, along with altruism, instant news sharing, socialization, self-promotion, and social media affordance, contribute to the circulation of fake news. In addition, Rubin [95] has proposed a conceptual model that identifies three minimal causal factors - automation, education, and regulation - that work together to facilitate the spread of fake news (epidemics) at the societal level, has suggested that information literacy efforts require interdisciplinary collaboration beyond library and information science, including media studies, journalism, psychology, and communication. Meanwhile, from the information behavior perspective, the information quality and credibility of the source can influence the perceived credibility of information [97]. The impact of information quality can be stronger than that of the source in some cases.

The final category of this theme focuses on sentiment analysis. Sentiment is one of the important aspects of information. Sentiment analysis is a technique of determining the sentiment of a subject, idea, or event from the content shared on social media by using natural language processing or computational linguistics techniques [96].

*Cluster 4 (yellow color): Hybrid-based as an intervention strategy.*
The fourth cluster, which is "hybrid-based", includes 9 pieces of literature. These pieces combine the criteria of the previous three clusters (cognition-based, automated-based, and information-based) and aim to address the spread of misinformation on social media by using one or more strategies, particularly in a domain specific to healthcare.

The spread of misinformation regarding health issues, especially during the COVID-19 pandemic, is a critical problem on social media. This has negatively impacted people's beliefs and contributed to vaccine hesitancy. Although research on how to tackle this problem is limited, it is crucial to adopt a comprehensive approach to combat it. This involves taking measures to restrict the spread of misinformation and create effective counter-messages. A study conducted by Islam et al. [9] has identified a variety of rumors and conspiracy theories that could erode people's confidence in the COVID-19 vaccine. The authors have suggested that policymakers should employ traditional verification approaches such as community engagement and risk communication, as well as establish evidence-based communication strategies, to address misinformation and potential vaccine disruptions. Effective policies play a crucial role in combatting misinformation on social media platforms, addressing data privacy violations, and mitigating the spread of fake news to create a more secure online environment [50], [98].

Furthermore, technologies such as topic modeling, AI, and machine learning can track and analyze vast amounts of media data in real-time. According to Sear et al. [55], using machine learning can overcome the scalability limitations of manual content analysis. Machine learning algorithms have the incredible ability to analyze various forms of content such as text, images, and videos to detect and categorize misinformation accurately. By examining intricate patterns within the content, these algorithms are proficient at identifying and flagging potentially false or misleading information.

The excessive use of social media has emerged as a significant public health issue. By examining how health misinformation is disseminated and its impact on people's attitudes, convictions, and actions, researchers can create effective intervention measures to combat its propagation. Pagoto et al. [53] proposed a public health agenda for social media research, which outlines ways to optimize social media usage for maximum health and wellness benefits while minimizing associated risks.

Educating people about false information is crucial. Vraga et al. [99] conducted a study that found combining news literacy messages with corrective responses effectively addressed health misinformation on Twitter. Correcting misinformation decreased its credibility and corrected misconceptions. However, exposure to misinformation lowered perceptions of news literacy without any boost from news literacy messages. Despite increased efforts to combat misinformation on social media, there remains significant uncertainty about intervention effectiveness.

Recent meta-analysis by Walter et al. [75] has introduced theory-driven moderators to clarify the effectiveness of social media interventions aimed at correcting health-related misinformation. The meta-analysis findings offer recommendations for combating health misinformation on social media.

### *I. COMPARATIVE ANALYSIS*

This subsection compares the themes identified in the previous subsection across various factors. Specifically, it will explore how the theme of cognition-based, automated-based, information-based, and hybrid-based is addressed in different contexts. Table VIII summarizes the results of this comparative analysis, providing a description, implementation examples, focus domain, publication output by journal, publication output by country, and high citation articles. By examining these factors, the analysis aims to highlight similarities and differences in approaches and offer valuable insights into global research trends and scholarly impact. The comparison discusses the following factors:

1. Description: provide a brief description of the theme.
2. Implementation example: Example of implementation and studies related to the theme.
3. Focus Domain: Domain focus of the theme.
4. Publication output by journal: The top 10 journals that ranked to be most productive in this field are mapped to the theme.
5. Publication output by country: The top 10 countries that ranked to be most productive in this field are mapped to the theme.
6. High citation articles: The high citation articles are mapped to the theme.

The similarity of cognition-based, automated-based, and information-based strategies is that they can be classified under domain-neutral clusters. However, each cluster leverages different techniques and approaches as intervention strategies to mitigate misinformation sharing. Cognition-based strategies focus on enhancing public awareness and critical thinking skills through educational campaigns, psychological interventions, and community engagement. Meanwhile, automated-based strategies leverage technologies such as algorithms, and fact-checking, as well as complying with social media platform policy to automatically detect and flag misinformation, employing an advanced content moderation system whereas information-based strategies explore and analyze information elements to identify the main reasons for content becoming misinformation that led to the sharing intention. Hybrid-based strategies combine the three criteria (cognition-based, automated-based, and information-based), to address and combat the spread of misinformation by combining one or more strategies, particularly in a domain specific to healthcare.

The comparative analysis of publication output by journals has revealed that the hybrid-based category is the most prevalent among the top productive journals in the field of misinformation intervention strategies. The Journal of Medical Internet Research has the highest number of publications on hybrid-based strategies, indicating a strong interdisciplinary approach to addressing health-related misinformation. Other journals like Computers in Human Behavior and New Media & Society also contribute significantly to hybrid-based strategies. Meanwhile, cognition-based was noted as the second chosen among the top productive journals, followed by information-based and lastly automated-based. Journals like Scientific Reports,

Information Systems Frontiers, Computers in Human Behavior, and Psychological Science have significant contributions to cognition-based strategies. The focus on cognition-based strategies indicates an emphasis on educational, psychological, and behavioral interventions. Furthermore, the Journal of Medical Internet Research, Social Media + Society, Internet Research, and New Media & Society have substantial publications in information-based strategies. Information-based strategies focus on improving the quality and accessibility of information to counter misinformation. In addition, journals such as Journal of Medical Internet Research, Social Media + Society, New Media & Society, and Digital Journalism have notable publications in automated-based strategies. This reflects a substantial interest in leveraging technology and automated systems to combat misinformation.

Publication output by country has shown that the USA has the highest number of publications across all categories of intervention strategies, particularly in cognition-based and hybrid-based strategies. This could be due to the significant research funding and resources available in the USA, as well as the strong focus on interdisciplinary approaches to combat misinformation. The UK shows a strong focus on cognition-based strategies, but there are no publications in the automated-based category. This might reflect a preference or greater expertise in educational and awareness-raising approaches within the UK. Meanwhile, the People's Republic of China has a more balanced approach with publications spread across all categories, although the numbers are generally lower compared to the USA. This might indicate a growing interest and investment in diverse intervention strategies to address misinformation.

The high-citation articles indicate that cognition-based and automated-based strategies are particularly influential in the field of misinformation intervention. The strong focus on health-related misinformation, especially during the COVID-19 pandemic, underscores the critical need for effective intervention strategies in this domain. The diverse approaches, including cognitive, automated, information-based, and hybrid strategies, highlight the multifaceted nature of combating misinformation and the importance of interdisciplinary research in this area.

TABLE VIII
COMPARATIVE ANALYSIS OF TYPOLOGY

| Criteria | Cognition-based | Automated-based | Information-based | Hybrid-based |
|---|---|---|---|---|
| Description | Cognition-based intervention strategies for mitigating misinformation sharing are designed to address the cognitive processes involved in how people interpret, retain, and share information. This theme provides works of literature that focus on nudging techniques, media literacy, and inoculation theory. | Automated-based intervention strategies for mitigating misinformation sharing leverage technology such as artificial intelligence and algorithm in mitigating the spread of misinformation. The literature on this theme focuses on four main objectives: identifying fake news through various means, using algorithms to detect fake news, employing fact-checking approaches, and developing policies and regulations for addressing fake news at the platform level. | Information-based intervention strategies for mitigating misinformation sharing focus on providing accurate information to educate the public, involving predictive modeling and analysis of factors influencing information sharing. The literature on this theme focuses on: information elements, constructing models, and sentiment analysis. | Hybrid-based intervention strategies for mitigating misinformation sharing combine the previous three criteria (cognition-based, automated-based, and information-based). This multifaceted strategy aims to address and combat the spread of misinformation by combining one or more strategies, particularly in a domain specific to healthcare. |
| Implementation Examples | • Nudging user using warnings or reminders that prompt users to consider for accuracy before sharing content.<br>• Campaign to promote media literacy by discuss strategies for recognizing misinformation<br>• "Bad News Game" help to build resilience against common tactics used in the spread of misinformation | • Machine learning algorithms are used to detect misinformation and collaborate with third-party fact-checkers to review content and provide corrections.<br>• Implementing a policy to regulate the sharing of content on social media platform. | • Interaction model of online information behaviors theorizing relationships among online information scanning, misinformation exposure, elaboration, sharing, and avoidance.<br>• Theoretical models that predict the sharing of misinformation can be a guideline for any practitioner.<br>• Interdisciplinary collaboration for information literacy involves media studies, journalism, psychology, and communication, where source quality | • Campaign public health agenda outlining ways to maximize social media for health benefits and minimize associated risks.<br>• The excessive use of social media has become a significant public health issue, and detection using machine learning can overcome the scalability limitations of manual content analysis.<br>• Combine news literacy messages with corrective |

|  |  |  |  | and credibility play a crucial role in influencing perceived information credibility. | responses effectively addressed health misinformation on Twitter. |
|---|---|---|---|---|---|
| Domain |  | Applicable to any domain | Applicable to any domain | Applicable to any domain | Applicable to health domain |

| | Publication output by journals | | | |
|---|---|---|---|---|
| | Cognition-based | Automated-based | Information-based | Hybrid-based |
| Journal of Medical Internet Research | 0 | 2 | 2 | 7 |
| Scientific Reports | 2 | 0 | 0 | 1 |
| Social Media + Society | 1 | 2 | 3 | 2 |
| Internet Research | 1 | 0 | 2 | 0 |
| Information Systems Frontiers | 2 | 1 | 0 | 2 |
| New Media & Society | 0 | 2 | 2 | 2 |
| Digital Journalism | 1 | 2 | 0 | 0 |
| Computers in Human Behavior | 2 | 0 | 1 | 3 |
| Information And Learning Sciences | 2 | 0 | 0 | 0 |
| Psychological Science | 2 | 0 | 0 | 1 |
| Journal of Medical Internet Research | 0 | 2 | 2 | 7 |
| Scientific Reports | 2 | 0 | 0 | 1 |

| | Publication output by country | | | |
|---|---|---|---|---|
| | Cognition-based | Automated-based | Information-based | Hybrid-based |
| USA | 31 | 22 | 17 | 28 |
| UK | 16 | 0 | 3 | 5 |
| People's Republic of China | 10 | 4 | 7 | 9 |
| Nigeria | 6 | 0 | 4 | 3 |
| Canada | 7 | 1 | 2 | 4 |
| Germany | 4 | 2 | 3 | 3 |
| Malaysia | 5 | 0 | 3 | 3 |
| Australia | 3 | 2 | 1 | 2 |
| Italy | 1 | 2 | 0 | 2 |
| Netherlands | 3 | 1 | 1 | 1 |

| | High citation articles | | | |
|---|---|---|---|---|
| | Cognition-based | Automated-based | Information-based | Hybrid-based |
| Fighting COVID-19 Misinformation on Social Media: Experimental Evidence for a Scalable Accuracy-Nudge Intervention | ✓ | | | ✓ |
| Healthcare information on YouTube: A systematic review | | | | ✓ |
| A Survey of Fake News: Fundamental Theories, Detection Methods, and Opportunities | | ✓ | | |
| Shifting attention to accuracy can reduce misinformation online | ✓ | | | |
| COVID-19 vaccine rumors and conspiracy theories: The need for cognitive inoculation against misinformation to improve vaccine adherence | ✓ | | | ✓ |
| Combating Fake News: A Survey on Identification and Mitigation Techniques | ✓ | ✓ | ✓ | |
| The Implied Truth Effect: Attaching Warnings to a Subset of Fake News Headlines Increases Perceived Accuracy of Headlines Without Warnings | ✓ | | | |
| A digital media literacy intervention increases discernment between mainstream and false news in the United States and India | ✓ | | | |
| Why Students Share Misinformation on Social Media: Motivation, Gender, and Study-level Differences | | | ✓ | |
| Evaluating the Impact of Attempts to Correct Health Misinformation on Social Media: A Meta-Analysis | | | | ✓ |

## V. DISCUSSION

In this section, we elaborated and provided the answers to research questions.

### A. RQ1: WHAT ARE THE CURRENT TRENDS IN INTERVENTION STRATEGIES FOR MISINFORMATION-SHARING STUDIES FROM 2013 TO 2023?

The structured identification of 139 relevant articles was made easier through the use of the PRISMA method, enabling a comprehensive analysis of trends over the past decade. Intervention strategies for addressing misinformation have undergone significant development over the course of a decade, spanning from 2013-2023. This evolution involved the transition from KB strategies (2015-2018) to TB strategies (2019-2023).

During the earlier years, the KB approach places a high priority on promoting information literacy. Several strategies have been employed to accomplish this goal, including regulatory enforcement, instructional efforts, and motivational programs that control the dissemination of false material on social media platforms. Strategies under the KB aim to raise public awareness of misinformation by promoting information literacy, educational campaigns, motivational initiatives, and the enforcement of policies regulating the spread of misinformation on social media platforms.

In recent years, intervention studies on misinformation have shifted towards taking TB approaches. TB techniques use technological advancements like machine learning and artificial intelligence (AI) to detect and minimize misinformation as a way to address the issues efficiently. Additionally, TB techniques for addressing misinformation-sharing were also focused on developing frameworks or tools as intervention strategies. Examples of these tools include flagging, warning, and nudging techniques to intervene in people's decision-making process. Social media platforms collaborate with third-party fact-checking organizations and crowdsource reports to flag suspicious content. These tools were embedded and integrated using algorithms into social media platforms.

The transition from using KB to TB approaches represents a broader trend in the field. This reflects a growing focus on integrating human behavior with technological advancements in research. By combining these approaches, the goal is to develop more robust and effective intervention strategies that can adapt to new forms of misinformation. Blending traditional KB methods with cutting-edge TB approaches allows stakeholders to develop a multifaceted and powerful strategy to tackle the pervasive issue of misinformation.

### B. RQ2: WHAT ARE THE MOST INFLUENTIAL COUNTRIES, JOURNALS, AND ARTICLES FOR STUDYING INTERVENTION STRATEGIES AGAINST MISINFORMATION-SHARING TOPICS?

This analysis of scholarly publications has identified the most influential countries, journals, and articles in the area of intervention strategies against misinformation-sharing. The findings revealed that the USA has become the top country in publishing articles related to this topic, followed by the UK, the People's Republic of China, Nigeria, and Canada. These countries were ranked as the top 5 countries with the most productive publications in intervention strategies for misinformation-sharing topics. The USA had far more publications than the second-place UK, likely due to its major concern about the spread of misinformation online. In addition, it is also possible that a high percentage of publications in the USA as most social media platforms are owned by them such as Meta. Furthermore, this is attributed to the manipulation of misinformation in many political events, providing the best breeding context and sufficient cases for the USA to study intervention strategies for misinformation sharing, such as the 2016 US presidential election. Furthermore, the USA excels in all categories of strategies such as cognition-based, automated-based, information-based, and hybrid-based strategies. On the other hand, the UK particularly stands out in cognition-based strategies. Meanwhile, the People's Republic of China demonstrates a more balanced approach across different strategies. The analysis indicates that the content of interest publications is significantly influenced by regional strengths and preferences.

Findings reveal that the Journal of Medical Internet Research is the most productive in this field. This study suggests that most misinformation studies relate to medical and health issues, as reflected by the top journals in the field. Additionally, the analysis of published works reveals that different journals make unique contributions to various approaches to combating misinformation. For instance, the Journal of Medical Internet Research extensively focuses on health-related issues, which allows it to dominate hybrid-based strategies. Journals such as Scientific Reports and Information Systems Frontiers contribute significantly to cognition-based strategies, while Social Media + Society and New Media & Society have a balanced focus across multiple strategies. The findings emphasize key journals and their contributions, guiding researchers and policymakers in identifying relevant sources and strategies for future work in this critical area.

The analysis of highly cited articles highlights distinct trends and focuses on the field of misinformation intervention. The article "Fighting COVID-19 Misinformation on Social Media: Experimental Evidence for a Scalable Accuracy-Nudge Intervention" was recognized as the most cited paper in this field. It demonstrated the importance of cognition-based and health-related issues in this area. The accumulative total of citations for the top 10 highly cited articles also highlights the significance of cognition-based and hybrid-based themes, positioning them at the top of the ranking. The high citation counts of these articles indicate the significant impact and recognition of these strategies within the academic community, guiding future research and policy-making in the fight against misinformation.

In conclusion, the study highlights the USA as the most influential country in the field, with the highest number of publications across all categories. The Journal of Medical Internet Research is the leading journal, especially in hybrid-based strategies. The article "Fighting COVID-19 Misinformation on Social Media: Experimental Evidence for a Scalable Accuracy-Nudge Intervention" stands out as the most cited and influential article, showcasing the most-wanted reference in this field of cognition-based and hybrid-based theme interventions. These insights emphasize the importance of interdisciplinary research and the integration of various strategies to effectively combat misinformation.

### C. RQ3: WHAT ARE THE EVOLVING THEMES AND TYPOLOGY FOR STUDYING INTERVENTION STRATEGIES AGAINST MISINFORMATION-SHARING TOPICS?

The typology of intervention strategies to combat misinformation on social media was generated based on findings revealed from keyword co-occurrence and bibliographic coupling analysis (see Fig. 5). This study has identified 4 main theme clustering: 1) cognition-based, 2) automated-based, 3) information-based, and 4) hybrid-based. The four clusters of strategies produce different approaches to combat misinformation sharing. The typology was categorized based on domain. Cluster 1-3 were classified under domain-neutral, while Cluster 4 was classified under health domain.

Intervention strategies within cluster 1 are cognition-based, designed to consider the cognitive processes involved in the decision to share information, including retention and interpretation. The approach aims to enhance cognitive abilities as a preventive measure against misinformation sharing such as employing nudging techniques, media literacy, and inoculation approaches.

Meanwhile, cluster 2 harnesses technology to counter the dissemination of misinformation through automated-based intervention strategies. It depends on technological solutions such as algorithms, fact-checking, and policy, which can be customized to detect and address misinformation at the platform level. The automated technique is both practical and well-suited for managing large volumes of information at the platform level. In addition, it is crucial to integrate platform policies into the algorithm to regulate users' activities on social media, particularly to combat the dissemination of misinformation. This adaptability makes these strategies effective across various domains.

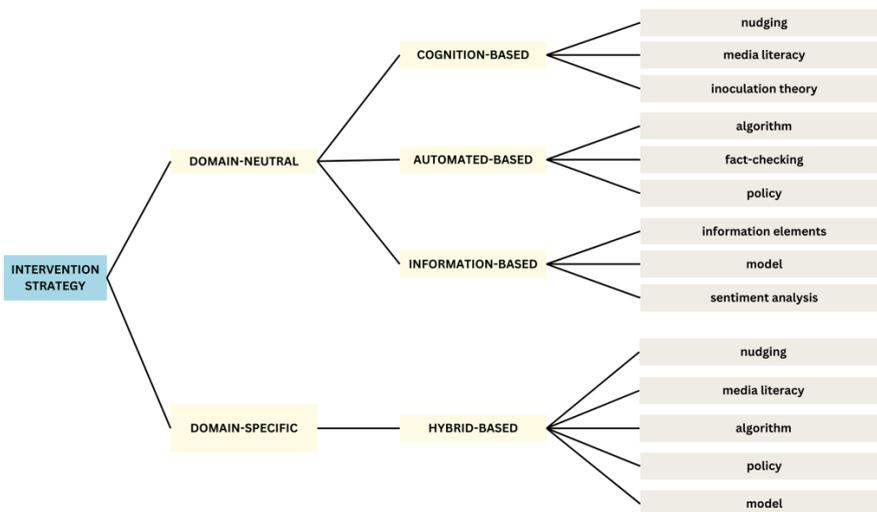

FIGURE 5. Typology of intervention strategy for misinformation sharing.

Furthermore, cluster 3 focuses on information-based interventions. This encompasses the dissemination of accurate information to educate the public, along with the utilization of predictive modeling and analysis to comprehend the factors influencing information sharing. The strategy uses information-based approaches like information elements (e.g.: information cues, characteristics, and literacy), models, and sentiment analysis to intervene in people's sharing behavior. Information features are analyzed and explored to identify the main reasons for content becoming misinformation that led to the sharing intention. These strategies are designed to enhance the quality and transparency of information regardless of the topic, thus applicable across multiple domains.

Hybrid-based (cluster 4) intervention strategies for mitigating misinformation sharing combine the previous three criteria (cognition-based, automated-based, and information-based). This multifaceted strategy aims to address and combat the spread of misinformation by combining one or more strategies, particularly in a domain-specific to healthcare. Hybrid-based strategies often address complex and critical issues, such as health misinformation, where the consequences of misinformation can be severe. By combining multiple approaches such as integrating educational efforts with AI-driven detection and fact-checking, these strategies offer a robust response tailored to the unique challenges of the health domain.

The comparative analysis findings indicate that intervention strategies have evolved into hybrid-based approaches, a theme that has been frequently published in the field's leading journal. These strategies are believed to be capable of addressing misinformation problems in today's advanced digital environment. Consequently, there is a need for intervention strategies that combine educational, technological, and informational approaches to effectively combat this issue. Healthcare publications are particularly focused on the hybrid-based theme due to its relevance in addressing real-time health crises, such as the COVID-19 pandemic and increasing public awareness of the dangers of health misinformation.

In addition, the cognition-based theme was noted as the most popular topic among high-citation articles in this field. The theme encompasses nudging techniques, media literacy, and inoculation approaches. The results of this study point to the necessity of using cognitive ability judgment to influence people's decision-making when it comes to sharing information because this issue has been widely researched. Moreover, the reasons for the popularity of cognition-based strategies include their empirical support, broad applicability, and long-standing foundation. The strategies are adaptable and work well in a range of situations to effectively mitigate misinformation. This demonstrates how crucial it is to improve public awareness and cognitive resistance to misinformation.

In conclusion, this study has introduced themes and typology for studying intervention strategies against misinformation topics. The findings highlight the importance of using a hybrid approach that is currently focused on the health domain but is also applicable across different domains. Additionally, cognition-based approaches were also important to increase public awareness and cognitive resistance for effectively combating misinformation, and they can also guide future research.

## VI. LIMITATIONS & CHALLENGES

Limitations were discovered during the review process. Despite providing valuable insights, the bibliometric analysis faced certain constraints in countering the spread of misinformation. Firstly, the analysis was limited to the WOS database, but future studies should consider utilizing additional databases for a more comprehensive review. Secondly, bibliometric reviews can be limited by misinterpretation and incomplete data. It may be necessary to read entire articles for a thorough analysis. Thirdly, there may be bias in the analysis due to variations in authors' names used in citations. For instance, some articles may have used initials while others included full middle names. Fourthly, the WOS database search using keywords may miss articles on intervention strategies against misinformation that use different phrases. This can result in incomplete search results. Finally, the data collection period was from the end of July-August 2023. Thus, to account for any potential increases in citations and publications, we analyzed the data based on the retrieval date. These limitations are typical of bibliometric studies conducted on any topic.

There are some challenges highlighted by several researchers in this field. Firstly, challenges in the intervention's design and length can persuade user's cognitive abilities [24], [61], [100]. It was very challenging to design intervention strategies which able to force users to 'think slow'. The design of the intervention was a challenge to attract user attention and make them think before sharing any misinformation. Studies on attention-based design are still lacking and provide a future research gap [24]. Secondly, the barriers to policies and regulation as intervention strategies for misinformation sharing on social media. Several research have investigated policies on social media platforms [10], [50], [98], [101]. Interference from many parties contribute to unclear policies and regulation by the practitioner (e.g.: government, social media platform) and contribute to barriers in designing the intervention strategies in the social media platform. Every country has different policies that make it tough to control the overload entry of information shared on social media. Improving the current regulation of social media platforms is also important to protect from the harms of misinformation on social media by controlling the source of the news from trusted and expert validation. Finally, research done by Gupta et al. [15] pinpointed various technical obstacles that must be overcome to effectively combat misinformation. These obstacles encompass a range of issues, including differences in defining what represents misinformation, the vast array of languages utilized by social media users globally, limitations in the quality and quantity of available datasets, and challenges associated with analyzing multimedia content.

## VII. CONCLUSION

This paper reviews and analyzes scholarly works related to strategies for combating the spread of misinformation on social media. The aim is to identify the themes and typology of interventions being used. The study has identified 4 important clusters (cognition-based, automated-based, information-based, and hybrid-based) and provides detailed insights into the strategies being employed and could be useful for future researchers in this field. The study used the PRISMA method to address research questions. This method systematically selects relevant studies, identifies trends, highlights influential countries, journals, and articles, and supports accurate theme development to enhance the credibility and relevance of findings on misinformation intervention strategies. The study has addressed three important questions regarding trends, key contributors (countries, journals, and articles), themes, and typology in the field. The results show that this topic is becoming progressively more popular, and this trend is probably going to continue. It is advised that countries collaborate to carry out comparative studies on the efficacy of hybrid-based intervention strategies across several domains to further enhance this field of study.

Comparative studies on intervention strategies against misinformation sharing across countries can provide broad guidelines, principles, and specific adaptations to enhance their effectiveness. This process can help expedite this process and ensure that strategies are adaptable to local situations through international cooperation. In addition to comparative studies, there is a need to focus on researching and developing hybrid technologies like combining strategies such as AI with nudging techniques, to combat misinformation across different domains. To solve this problem more effectively, it is critical to investigate the significance and efficacy of this strategy.

Meanwhile, understanding social media user behavior is also important for developing effective interventions. Comparative research should look at how different consumer demographics engage with misinformation and react to different intervention strategies. Additionally, understanding user behavior, can direct the creation of more tailored strategies that consider sharing behavior and psychological factors that connect with specific target audiences. Through collaboration efforts, future researchers can leverage diverse contexts and experiences in a variety of cultural and sociopolitical circumstances.

In conclusion, given the urgency that society needs to combat misinformation, more research into intervention measures aimed at reducing the spread of misinformation is crucial. To effectively address the complex and dynamic difficulties of misinformation in today's digital environment and eventually promote a more resilient and informed global community, it is imperative to design effective solutions that can be tailored to varied situations.

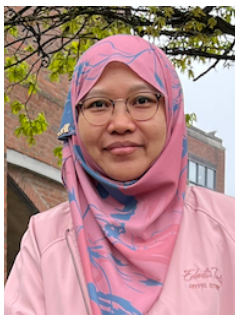

**JUANITA ZAINUDIN,** currently pursuing a Ph.D. degree with the Institute of Visual Informatics (IVI) at Universiti Kebangsaan Malaysia. Alongside her studies, she also works as a lecturer at the Faculty of Computing and Multimedia at Universiti Poly-Tech Malaysia where she imparts her knowledge to the next generation of scholars. Her research is focused on the field of Human Computer Interaction (HCI), information behavior, and misinformation sharing. She is a member of IEEE.

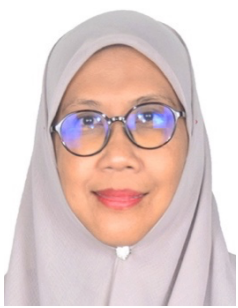

**NAZLENA MOHAMAD ALI** received the Ph.D. degree in Human Computer Interaction from Dublin City University, Ireland. She is currently an Associate Professor and a Senior Research Fellow with the Institute of Visual Informatics (IVI), Universiti Kebangsaan Malaysia. Her research interests include interaction design, UI/UX, persuasive technology, and usability.

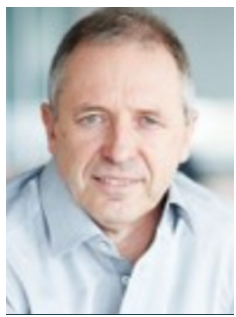

**ALAN F. SMEATON** is currently a Professor of computing and the Former Director of the Insight–Centre for Data Analytics, Dublin City University. His research interests include human memory, why we forget some things and not others, and how we can use technology like search systems, to compensate for when we do forget. He was the winner of the Royal Irish Academy Gold Medal for Engineering Sciences, in 2015. He is the Chair of ACM SIGMM and an IEEE Fellow.

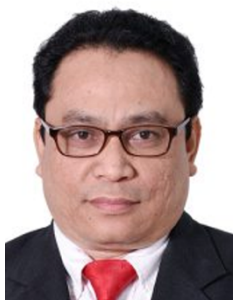

**MOHAMAD TAHA IJAB** is currently a research fellow at the Institute of Visual Informatics (IVI), Universiti Kebangsaan Malaysia. He received his Ph.D. degree from RMIT University, Australia in the field of Business IT. He is interested and researching in the fields of Social & Environmental Informatics, Business Process Management, and Organizational Management in Informatics.